\documentclass[11pt]{article}
\usepackage{amsmath,amssymb,color,graphics,epsfig,cite}
\usepackage{amsfonts}
\numberwithin{equation}{section}
\newcommand{\hoch}[1]{$\, ^{#1}$}

\newcommand{\be}{\begin{equation}}
\newcommand{\ee}{\end{equation}}
\newcommand{\bea}{\setlength\arraycolsep{2pt} \begin{eqnarray}}
\newcommand{\eea}{\end{eqnarray}}

\def\fft#1#2{{\frac{#1}{#2}}}

\def\0{{\sst{(0)}}}
\def\1{{\sst{(1)}}}
\def\2{{\sst{(2)}}}
\def\3{{\sst{(3)}}}
\def\4{{\sst{(4)}}}
\def\5{{\sst{(5)}}}
\def\6{{\sst{(6)}}}
\def\7{{\sst{(7)}}}
\def\8{{\sst{(8)}}}
\def\sst#1{{\scriptscriptstyle #1}}

\begin{document}
\begin{center}
{\Large {\bf Rotating Black Holes and the Kerr/CFT Correspondence in Einstein-Bumblebee Gravity}}

\vspace{20pt}

{\large Yu-Qi Chen\hoch{1}, Jin-Yang Shen\hoch{2} and Hai-Shan Liu\hoch{1}}

\vspace{10pt}

{\it \hoch{1}Center for Joint Quantum Studies and Department of Physics,\\
School of Science, Tianjin University, Tianjin 300350, China }

\vspace{5pt}

{\it \hoch{2}School of Physics and Electronics, Hunan University, Changsha 410082, China}

\vspace{40pt}

\underline{ABSTRACT}

\end{center}

We constructed rotating black holes with equal angular momentum in five dimensional Einstein-Bumblebee gravity with and without cosmological constant. Their thermodynamic properties are examined via two distinct methods: the Wald formalism and the Komar integral. Notably, the conserved charges, including mass, angular momentum, and entropy, computed from these two approaches differ by a constant prefactor that is solely determined by the Bumblebee coupling. Subsequently, we apply the Kerr/CFT correspondence to derive the microscopic entropy of these black holes and find that it precisely reproduces the entropy in Komar-integral version, rather than the Wald entropy.

\vfill{\footnotesize yuqi\_chen@tju.edu.cn ~~~ Jinyang\_Shen@hnu.edu.cn~~~~hsliu.zju@gmail.com(corresponding author) }
%\vfill {\footnotesize mrhonglu@gmail.com}

%{\footnotesize \hoch{*}Corresponding author}

\thispagestyle{empty}
\pagebreak

\tableofcontents
\addtocontents{toc}{\protect\setcounter{tocdepth}{2}}

\newpage

\section{Introduction}
For over a century, Einstein's general relativity has passed every rigorous experimental test with breathtaking precision, from perfectly explaining Mercury's anomalous orbit and predicting the bending of starlight during solar eclipses, to the modern-day detections of gravitational waves and the direct imaging of a black hole's shadow\cite{LIGOScientific:2017vwq,EventHorizonTelescope:2019dse}. However, the incompatibility of general relativity with quantum mechanics, together with mounting evidence from cosmology and large‑scale structure, strongly points to the need for extensions beyond Einstein’s framework. Among the various modified gravity proposals, theories that implement spontaneous Lorentz symmetry breaking have garnered particularly rapid attention over the past several decades. Lorentz invariance stands as a foundational pillar of both special and general relativity. Even so, numerous candidate quantum gravity constructions, like string theory, imply that this symmetry may hold only as a low-energy approximation, susceptible to violation at high energy scales. Investigations of Lorentz-violating effective theories accordingly offer a valuable avenue for probing novel physics lying outside the scope of standard Einstein gravity.In this context, the Einstein–Bumblebee theory provides one of the simplest and most controllable frameworks to explore spontaneous Lorentz symmetry breaking in a covariant setting\cite{Bluhm:2004ep,Kostelecky:1988zi,Kostelecky:2003fs}. In the Bumblebee theory, depending on whether the vacuum expectation value is timelike or spacelike, the resulting theory exhibits anisotropic modifications to the effective spacetime structure and to the propagation of gravitational perturbations\cite{Maluf:2014dpa,Amarilo:2023wpn}. The Einstein–Bumblebee model is therefore an ideal laboratory for studying how Lorentz symmetry breaking may affect gravitational waves\cite{Maluf:2014dpa,Amarilo:2023wpn,Liang:2022hxd,Liu:2022dcn,Liu:2026cxs}, quantum information\cite{Liu:2024oeq,Liu:2024wpa,Liu:2025bpp,Liu:2026vvp,AraujoFilho:2025nmc}, and black hole physics\cite{Casana:2017jkc,Ding:2019mal,Ding:2023niy,Maluf:2020kgf,Xu:2022frb,An:2024fzf,Chen:2025ypx}. 

Within the Einstein–Bumblebee framework, a distinct solution arises when the bumblebee vector field adopts a purely radial profile; in this configuration, the field diverges at the black hole event horizon. This behavior closely parallels that of Horndeski gravity black holes, where the scalar field similarly diverges on the horizon. It should be noted, however, that the Wald entropy formula for computing black hole entropy may become subtle in such cases \cite{Feng:2015oea,Feng:2015wvb}. A number of formalisms are available for calculating black hole thermodynamic quantities. Among them are the covariant phase space method\cite{An:2024fzf,Chen:2025ypx,Li:2025tcd}, the Komar integral method\cite{Ding:2022qcy}, and the ADM formalism \cite{Liu:2024axg}. Yet, investigations of Horndeski gravity black holes have shown that the resulting entropy expressions are method‑dependent, casting doubt on their uniqueness. In the context of Bumblebee gravity, the correct entropy definition remains similarly unsettled and awaits further scrutiny.
  
On the other hand, considerable efforts have been devoted to exploring the microscopic origin of black hole entropy. For instance, Strominger and Vafa famously showed that the Bekenstein–Hawking entropy for a class of five-dimensional extremal black holes in string theory can be derived by counting the degeneracy of BPS bound states \cite{Strominger:1996sh}.

In parallel, one of the most significant developments in theoretical physics over the past three decades has been the proposal of the AdS/CFT correspondence\cite{Maldacena:1997re,Witten:1998qj,Gubser:2002tv}, which provided a powerful framework for studying the microscopic structure of black holes \cite{Ryu:2006bv,Almheiri:2019hni,Penington:2019npb}. The AdS/CFT conjecture originates from the study of dualities in string theory, relating weakly coupled type IIB supergravity in \(\text{AdS}_5\times S^5\) spacetime to strongly coupled \(\mathcal{N}=4\) super Yang–Mills theory on its boundary. This duality establishes a profound connection between higher-dimensional gravity and codimension-1 quantum field theory and has achieved a series of remarkable results in strongly coupled systems \cite{Hartnoll:2008vx,Roberts:2008ns,Amado:2009ts,Montull:2009fe,Nishioka:2009zj,Cai:2010cv,Cai:2013aca,Basu:2008bh,Kruczenski:2003uq,Sakai:2004cn,Sakai:2005yt,Shen:2025yrn}. From a bottom-up perspective, the existence of a quantum field theory on the spacetime boundary implies, through holography, that the asymptotic symmetry group (ASG) of the bulk geometry must coincide with the conformal symmetry group of the boundary theory. The ASG is defined as a quotient group of allowed diffeomorphisms modulo the set of trivial diffeomorphisms, namely
\[
\text{ASG}:=\frac{\text{Allowed Symmetry Transformations}}{\text{Trivial Symmetry Transformations}}.
\]
A key observation is that the conserved charges associated with the diffeomorphisms of the three-dimensional BTZ black hole realize the quantum Virasoro algebra structure of the dual two-dimensional conformal field theory. As a consequence, the thermal entropy of the 2D CFT given by Cardy \cite{Cardy:1986ie} precisely matches the Bekenstein–Hawking entropy of the corresponding BTZ black hole, thereby holographically revealing the microscopic origin of black hole entropy \cite{Banados:1992wn}. Hence, the construction of the ASG offers a powerful approach to probe holographic principles beyond the standard AdS/CFT correspondence \cite{Guica:2008mu,Barnich:2010eb} and to elucidate the microscopic origin of entropy in black holes that are not asymptotically AdS. 

Among these, a particularly elegant realization of holography in asymptotically flat spacetimes is provided by the Kerr/CFT correspondence, first proposed by Guica, Hartman, Song, and Strominger\cite{Guica:2008mu}. This correspondence associates the near-horizon geometry of an extremal Kerr black hole—known as the NHEK geometry—with a chiral two-dimensional CFT. The Cardy formula for the CFT entropy again precisely reproduces the Bekenstein–Hawking entropy of the extremal Kerr black hole. This success provides strong evidence that the microscopic degrees of freedom responsible for Ricci-flat black hole entropy can be described by a dual conformal field theory. Subsequently, this result was extended to non-extremal Kerr black holes \cite{Castro:2010fd} and higher-dimensional cases\cite{Lu:2008jk}. In some later extensions, \cite{Krishnan:2009tj,Liu:2021hvb} suggested that the generalization of the Kerr/CFT correspondence to Einstein gravity extended with an \(R_{\mu\nu\rho\sigma}R^{\mu\nu\rho\sigma}\) term is subtle. Meanwhile, \cite{Nian:2020iqh} proposed that the Kerr/CFT correspondence could be invoked to resolve the black hole information loss paradox. Overall, the Kerr/CFT correspondence can serve as a probe for investigating the microscopic entropy of black holes in extended gravity theories. Therefore, it offers a promising framework for probing the microscopic origin of black hole entropy in Bumblebee gravity.

The Kerr/CFT correspondence is intrinsically designed for rotating black holes. Although it offers a powerful tool for probing their microscopic entropy, its application to Einstein--Bumblebee gravity faces an immediate obstacle: the construction of rotating black hole solutions within this framework remains scarce and largely unsatisfactory. The first exact rotating solution, referred to as the Kerr-like black hole, was constructed in \cite{Ding:2019mal}. However, as noted in \cite{Maluf:2022knd}, this solution does not actually satisfy the bumblebee field equations. Subsequent studies attempted to construct exact four-dimensional rotating black hole solutions by introducing an additional \(\theta\)-component of the bumblebee field \cite{Poulis:2021nqh,Ovcharenko:2026rvj}; the trade-off, however, is that this introduces extra terms into the \(\theta,r\)-components of the metric, rendering the analysis considerably more intricate. A rotating black hole with a timelike bumblebee field was also constructed in\cite{Xu:2026zgd}. To date, no further rotating black hole solutions have been constructed within this framework. Given this state of affairs, the present work pursues two main objectives:
\begin{itemize}
	\item The construction of novel rotating black hole solutions within the Bumblebee framework.
	\item The computation of their microscopic entropy through the Kerr/CFT correspondence, thereby testing the agreement between different entropy formalisms.
\end{itemize}

The paper is organized as follows. Section 2 provides a review of Einstein--Bumblebee gravity and the covariant phase space formalism. Section 3 is devoted to the construction of a five-dimensional rotating black hole solution with equal angular momenta, along with a thorough analysis of its thermodynamic properties. In Section 4, we apply the Kerr/CFT correspondence to investigate the holographic description and microscopic entropy of these higher-dimensional rotating black holes. Finally, Section 5 contains our discussion and concluding remarks.

\section{Bumblebee Gravity}
\subsection{Theory}
One of the simplest toy-model of the gravity theory with spontaneous Lorentz symmetry breaking is the bumblebee gravity which is proposed by Kostelecky\cite{Bluhm:2004ep,Kostelecky:1988zi,Kostelecky:2003fs}. It is a typical vector-tensor theory described by a vector field  $B^{\mu}$ that couples non-minimally to gravity via the Ricci tensor $R_{\mu\nu}$. The whole action of the bumblebee gravity is given by \cite{Casana:2017jkc}
\begin{equation}\label{action}
    I=\int d^{D}x\sqrt{-g}[\frac{1}{2\kappa}(R-2\Lambda+\gamma B^{\mu}B^{\nu}R_{\mu\nu})-\frac{1}{4}B_{\mu\nu}B^{\mu\nu}-V(B^{\mu})],
\end{equation}
where $\gamma$ is a real constant that describes the non-minimal coupling, $\kappa$ is chosen as $8\pi$ and $B_{\mu\nu}$ is the bumblebee field strength given by $B_{\mu\nu}=\partial_{\mu}B_{\nu}-\partial_{\nu}B_{\mu}$. The Lorentz symmetry breaking originates from the Higgs mechanism, which requires a nonzero vacuum expectation value (VEV) of the bumblebee field  $B_{\mu}$. The VEV of the bumblebee field is denoted as
\begin{equation}
    <B_{\mu}>=b_{\mu}.
\end{equation}
And, the bumblebee potential has the form 
\begin{equation}
    V(B_\mu)=V(X)~,~ \text{with}~X=B_{\mu}B^{\mu}\pm b^2,
\end{equation}
where $b^2$ represents a real positive constant. In this work, we consider the bumblebee field frozen in the vacuum state which implies the condition on the VEV of bumblebee field $ b^{\mu}b_{\mu}  \pm b^2 = 0 $. Thus  $b^{\mu}b_{\mu}=\mp b^2$, where $\mp$ signs mean that $b^{\mu}$ is timelike or spacelike, respectively.
The Einstein equations can be directly obtained through variation of the metric, which are given by 
\begin{equation}\label{EOM-gr}
    \begin{split}
        E_{\mu\nu}=&R_{\mu\nu}-\frac{1}{2}g_{\mu\nu}R+\Lambda g_{\mu\nu}-\kappa T_{\mu\nu}^{(Bee)} = 0,
    \end{split}
\end{equation}
with
\begin{equation}
    \begin{split}
        T_{\mu\nu}^{(Bee)}=&-b_{\mu\alpha}b^{\alpha}_{~\nu}-\frac{1}{4}b_{\alpha\beta}b^{\alpha\beta}-g_{\mu\nu}V+2b_{\mu}b_{\nu}V'+\frac{\gamma}{\kappa}(\frac{1}{2}b^{\alpha}b^{\beta}R_{\alpha\beta}g_{\mu\nu}-b_{\mu}b^{\alpha}R_{\alpha\nu}-b_{\nu}b^{\alpha}R_{\alpha\mu}\\&+\frac{1}{2}\nabla_{\alpha}\nabla_{\mu}(b^{\alpha} b_{\nu})
        +\frac{1}{2}\nabla_{\alpha}\nabla_{\nu}(b^{\alpha} b_{\mu})-\frac{1}{2}\nabla^2(b_{\mu}b_{\nu})-\frac{1}{2}g_{\mu\nu}\nabla_{\alpha}\nabla_{\beta}(b^{\alpha}b^{\beta})).
    \end{split}
\end{equation}
The "prime" of potential $V$ denotes the derivative of potential functional with respect to its argument, $V(X)'=dV(X)/dX$. And the equations of motion for bumblebee field are derived through the variation of the vector field $b_{\mu}$, namely
\begin{equation}\label{EOM-bee}
  E_{\nu}=\nabla^{\mu}b_{\mu\nu}+\frac{\gamma}{\kappa}b^{\mu}R_{\mu\nu}-2V'b_{\nu} =0.
\end{equation}

\subsection{Covariant Phase Space Formalism}

As noted in the introduction, we shall investigate the thermodynamics and central charge of rotating black holes within the framework of bumblebee gravity theory. A powerful tool we will employ to study both of these components is the Covariant Phase Space Method (CPSM)\cite{Wald:1993nt,Iyer:1994ys}.

CPSM is a framework designed to define conserved quantities in gravity theories with diffeomorphism invariance. Distinct from canonical Hamiltonian approaches, which require a $3+1$ decomposition of spacetime into space and time, CPSM is manifestly covariant and formulated entirely in terms of differential forms. Consider a D-dimensional spacetime manifold $\mathcal{M}$ with dynamical fields collectively denoted by $\Phi$, the Lagrangian is given by $\mathcal{L}(\Phi)$. If there exist a diffeomorphism $\xi$, the variation of the Lagrangian can be written as
\begin{equation}
    \delta_{\xi}\star \mathcal{L}(\Phi)=EOM\delta_{\xi}\phi+d\Theta(\phi,\delta_{\xi}\phi),
\end{equation}
where $\Theta(\phi,\delta_{\xi}\phi)$ is the symplectic potential, a (D-1)-form depending on the field and its variation. It captures the boundary contributions of the variation and serves as the foundation for constructing a symplectic structure. The symplectic current 2-form is thus defined as 
\begin{equation}
    \omega(\delta\Phi,\delta_{\xi}\Phi,\Phi)=\delta\Theta(\delta_{\xi}\Phi,\Phi)-\delta_{\xi}\Theta(\delta\Phi,\Phi).
\end{equation}
This current is a (D-1)-form on spacetime and a 2-form on the tangent space of field configurations. The current is closed when the field $\Phi$ satisfies the on-shell condition $EOM=0$,
\begin{equation}
    d\omega(\delta\Phi,\delta_{\xi}\Phi,\Phi)\approx 0.
\end{equation}
where $\approx$ denotes the on-shell equality. The symplectic current thus can be written as a total derivative of a ($D-2$) form $\omega\approx dk$. The total Hamiltonian of the diffeomorphism can be expressed as 
\begin{equation}
    \delta H=\frac{1}{16\pi}\int_{\mathcal{M}}\omega=\frac{1}{16\pi}\int_{\mathcal{M}} dk=\frac{1}{16\pi}\int_{\partial\mathcal{M}}k.
\end{equation}
Then to concretely construct the symplectic current, we introduce a closed ($D-1$) form $J$
\begin{equation}
    J=\Theta_{D-1}-i_{\xi}\star L~,~~dJ=0,
\end{equation}
where $i_{\xi}$ denotes contracting the subsequent tensor with $\xi$. It  can thus also be written as $J\approx dQ$. Upon taking the variation of the ($D-1$) form $J$, we find
\begin{equation}
    \delta J=\omega+di_{\xi}\Theta.
\end{equation}
It gives 
\begin{equation}
    \omega=d(\delta Q-i_{\xi}\Theta)=dk_{\xi}[\phi,\delta\phi],
\end{equation}
where
\begin{equation}
    k_{\xi}[\phi,\delta\phi]=\delta Q-i_{\xi}\Theta.
\end{equation}
As discussed in\cite{An:2024fzf,Chen:2025ypx}, in the framework of Einstein-bumblebee gravity, the surface term is
\begin{equation}\label{surface}
\begin{split}
    &\Theta^{\mu}=2P^{\mu\nu\rho\sigma}\nabla_{\sigma}\delta_{\xi} g_{\rho\nu}-2\nabla_{\nu}P^{\rho\mu\nu\sigma}\delta_{\xi} g_{\rho\sigma}-2\kappa B^{\mu\nu}\delta_{\xi} B_{\nu}.    
\end{split}
\end{equation}
where 
\begin{equation}\label{pabcd}
    \begin{split}
      P^{\mu\nu\rho\sigma}&=\frac{\partial \mathcal{L}}{\partial R_{\mu\nu\rho\sigma}}\\&=\frac{\gamma}{4}(B^{\nu}B^{\sigma}g^{\rho\mu}-B^{\mu}B^{\sigma}g^{\rho\nu}-B^{\rho}B^{\nu}g^{\mu\sigma}+B^{\rho}B^{\mu}g^{\nu\sigma})+\frac{1}{2}(g^{\mu\rho}g^{\nu\sigma}-g^{\nu\rho}g^{\mu\sigma}).  
    \end{split}
\end{equation}
And the Noether charge $Q^{\mu\nu}$ can be directly read-off from the on-shell relation $J\approx dQ$, namely
\begin{equation}\label{current}
\begin{split}
    J^{\mu}&=\Theta^{\mu}-\xi^{\mu}L
    \\&=\nabla_{\nu}(-2\nabla^{[\mu}\xi^{\nu]}+2\gamma(\xi_{\sigma}\nabla^{\mu}(B^{\nu}B^{\sigma})-\xi^{\mu}\nabla_{\sigma}(B^{\nu}B^{\sigma})-B^{\sigma}B^{\mu}\nabla_{\sigma}\xi^{\nu})-2\kappa B^{\mu\nu}B_{\alpha}\xi^{\alpha})
    \\&=\nabla_{\nu}Q^{\mu\nu}.
\end{split}
\end{equation}
Accordingly, the explicit form of the Noether charge $Q^{\mu\nu}$ is 
\begin{equation}\label{Komar}
\begin{split}   Q^{\mu\nu}&=-2\nabla^{[\mu}\xi^{\nu]}+2\gamma(\xi_{\sigma}\nabla^{\mu}(B^{\nu}B^{\sigma})-\xi^{\mu}\nabla_{\sigma}(B^{\nu}B^{\sigma})-B^{\sigma}B^{\mu}\nabla_{\sigma}\xi^{\nu})-2\kappa B^{\mu\nu}B_{\alpha}\xi^{\alpha}.
\end{split}
\end{equation}
We denote the deviation of metric as $\delta g_{\mu\nu}=h_{\mu\nu}$. The symplectic form $k$ can be written as
\begin{equation}\label{k-bee}
    k_{\xi}[h,g]=\sqrt{-g}\epsilon_{a_{1}...a_{n-2}\mu\nu}(\delta Q^{\mu\nu} -\xi^{\nu}\Theta^{\mu}).
\end{equation}
where the $\delta Q^{\mu\nu}$ is defined by
\begin{equation}
\begin{split}
\delta Q^{\mu\nu}&=\frac{1}{2\sqrt{-g}}\delta (\sqrt{-g}Q^{\mu\nu})\\&=-\frac{1}{4}hQ^{\mu\nu}+\Delta_{1}+\gamma(\Delta_{2}+\Delta_{3}+\Delta_{4})+\Delta_{5},
\end{split}
\end{equation}
with
\begin{equation}
    \begin{split}
        &\Delta_{1}=-(-h^{\mu\alpha}\nabla_{\alpha}\xi^{\nu}+\delta\Gamma^{\nu}_{\alpha\beta}\xi^{\beta}g^{\mu\alpha}),
        \\&\Delta_{2}=\xi^{\sigma}[-\nabla_{\alpha}(B^{\nu}B_{\sigma})h^{\mu\alpha}+\nabla^{\mu}\delta(B_{\sigma }B^{\nu})+\delta\Gamma^{\nu}_{\alpha\beta}B^{\beta}B_{\sigma}g^{\mu\alpha}-\delta\Gamma^{\beta}_{\alpha\sigma}B^{\nu}B_{\beta}g^{\mu\alpha},
        \\&\Delta_{3}=-\xi^{\mu}[\nabla_{\sigma}\delta(B^{\nu}B^{\sigma})+\delta\Gamma^{\nu}_{\sigma\alpha}B^{\alpha}B^{\sigma}+\delta\Gamma^{\sigma}_{\sigma\alpha}B^{\nu}B^{\alpha}],
        \\&\Delta_{4}=-B^{\mu}\nabla_{\sigma}\xi^{\nu}\delta B^{\sigma}-B^{\sigma}\nabla_{\sigma}\xi^{\nu}\delta B^{\mu}-B^{\sigma}B^{\mu}\xi^{\alpha}\delta\Gamma^{\nu}_{\sigma\alpha},\\&\Delta_{5}=-\kappa(2h^{[\mu}_{~\sigma}B^{\nu]\sigma}B_{\alpha}\xi^{\alpha}+B^{\mu\nu}\xi^{\alpha}\delta B_{\alpha}+2g^{\mu[\rho}g^{\sigma]\nu}B_{\alpha}\xi^{\alpha}\partial_{\rho}\delta B_{\sigma}),
    \end{split}
\end{equation}
where
\begin{equation}
\begin{split}
     &\delta\Gamma^{\alpha}_{\mu\nu}=\frac{1}{2}g^{\alpha\beta}(\nabla_{\mu}h_{\nu\beta}+\nabla_{\nu}h_{\mu\beta}-\nabla_{\beta}h_{\mu\nu}),\\&\delta B^{\mu}=g^{\mu\nu}\delta B_{\nu}-B_{\nu}h^{\mu\nu}.
\end{split}
\end{equation}
Covariant Phase Space Formalism indeed provides a systematic framework for analyzing conserved charges associated with transformation described by a given vector $\xi$.

\section{Rotating Black Hole with Equal Angular Momentum}

Compared to static black hole solutions, the expressions for rotating black holes are considerably more complex. Consequently, rotating black holes are difficult to obtain by solving Einstein’s equations. In this section, we attempt to construct a special class of rotating black holes, those with equal angular momentum. We begin with the simple case of five-dimensional Einstein-Bumblebee gravity without a cosmological constant.

\subsection{Black hole solution in Bumblebee gravity without cosmological constant}
In Einstein gravity, the five-dimensional Myers–Perry black hole offers a natural generalization of the Kerr solution and possesses two independent angular momenta\cite{Myers:1986un}. When the two rotation parameters are set equal, the solution simplifies considerably, and the metric reduces to cohomogeneity-one form. Under this restriction, the metric reads
\be
ds^2_\5  = \fft{dr^2}{h(r)} - \fft{h(r)}{W(r)} dt^2 + \fft 14 r^2 W(r) ( d\psi + \cos\theta d \phi + \omega(r) dt )^2 + \fft 14 r^2 (d\theta^2 + \sin^2 \theta d\phi^2 ) \,.
\ee
where
\be
h(r)= 1 - \fft{2 \mu}{r^2} + \fft{\nu^2}{r^4} \,, \quad W(r)=1+\fft{\nu^2}{r^4} \,, \quad \omega(r) = -\fft{2 \sqrt{2 \mu} \nu}{  (r^4+\nu^2)} \,.
\ee
This metric can also describe a series of other physical models in different parameter spaces, such as solitons or time machines\cite{Feng:2016dbw}. In this paper, we focus solely on the black hole solution with the parameter choice $(\mu>0,\nu\in\mathbb{R})$. 

Motivated by this structure, we adopt the following ansatz in Einstein–bumblebee gravity:
\be\label{ansatz}
ds^2_\5  = \fft{dr^2}{f(r)} - \fft{h(r)}{W(r)} dt^2 + \fft 14 r^2 W(r) ( d\psi + \cos\theta d \phi + \omega(r) dt )^2 + \fft 14 r^2 (d\theta^2 + \sin^2 \theta d\phi^2 ) \,,
\ee
where $h(r),f(r),W(r),\omega(r)$ are unknown radial functions to be determined. We take the bumblebee field to possess a spacelike vacuum expectation value oriented purely along the radial direction, with
\begin{equation}\label{ansatz-bee}
    b_{\mu}=(0,b(r),0,0,0).
\end{equation}
Since the non-zero expectation value condition $b^{\mu}b_{\mu}=b^2=constant$ satisfies, the profile $b(r)$ of the bumblebee field can be directly solved as
\begin{equation}
    b^{\mu}b_{\mu}=g^{rr}b(r)^2=b^2~\rightarrow ~b(r)=\sqrt{\frac{b^2}{f(r)}}.
\end{equation}
Then we consider a vector potential with high power
\begin{equation}
    V(X)=\frac{\lambda}{2}X^n~,~~n\geq 2,
\end{equation}
where $X=B^{\mu}B_{\mu}\pm b^2$. As a consequence, the potential and its first derivative vanish identically on the vacuum state, namely
\begin{equation}
	\begin{split}
		&V(B_{\mu}B^{\mu}\pm b^2)\Big|_{B^{\mu}=b^{\mu}}=0,
		\\&V'(B_{\mu}B^{\mu}\pm b^2)\Big|_{B^{\mu}=b^{\mu}}=0.
	\end{split}
\end{equation}
This indicates that the potential terms make no contribution to the equations of motion. Substituting the ansatz in Eq.\eqref{ansatz} into the field equations $(E_{\mu\nu}=0,E_{\nu}=0)$, one can obtain a specific black hole solution, namely
\be
f(r)=\fft{h(r)}{1+l}\,,\quad h(r)= 1 - \fft{2 \mu}{r^2} + \fft{\nu^2}{r^4} \,, \quad W(r)=1+\fft{\nu^2}{r^4} \,, \quad \omega(r) = -\fft{2 \sqrt{2 \mu} \nu}{  (r^4+\nu^2)} \,,
\ee
where the constant $l$ is the Lorentz symmetry breaking factor defined as $l=\gamma b^2$. The metric functions $h,W,\omega$ are consistent with that of Einstein's case, but the function $f$ is scaled by the factor $1/(l+1)$. This solution possesses two horizons, located at $r_{\pm}$, which the real root of $h(r)=0$, namely
\begin{equation}
    r_{\pm}=\sqrt{\mu\pm\sqrt{\mu^2-\nu^2}}.
\end{equation}
To illustrate that our solution is indeed non-trivial compared to the original Myers-Perry black hole, we compute the Kretschmann scalar, a key invariant for characterizing spacetime curvature, which is given by
\begin{equation}
    R_{\mu\nu\rho\sigma}R^{\mu\nu\rho\sigma}=\frac{4[(96+80l+11l^2)\nu^4+3r^4(24\mu^2+4l\mu r^2+l^2r^4)-2\nu^2((96+58l)\mu r^2+l^2r^4)]}{(1+l)^2r^{12}}.
\end{equation}
when $l\rightarrow 0$, the solution reduces to that of Einstein gravity. Our result is a complicated function of $l$, which implies that our new solution cannot be recovered from the Myers-Perry solution via a coordinate transformation. Another important piece of evidence distinguishing our solution from the Myers-Perry black hole is that the Ricci scalar,
\begin{equation}
    R=-\frac{2l(\nu^2-3r^4)}{(1+l)r^6},
\end{equation}
is unequal to zero. It indicates that our new rotating black hole is no longer Ricci-flat.

%\textcolor{red}{In Einstein--Bumblebee gravity, the first exact rotating solution—referred to as the Kerr-like black hole—was constructed in [XXX]. However, as pointed out in [XXX], this solution fails to satisfy the bumblebee field equations, with \(E_{\nu}\neq 0\). To address this issue, subsequent work introduced an additional \(\theta\)-component of the bumblebee field in an attempt to construct exact four-dimensional rotating black holes [XXX]. This comes at the cost of introducing extra terms into the \(\theta,r\)-components of the metric, which significantly complicates the analysis. Moreover, no higher-dimensional rotating black hole solutions have yet been constructed within this framework. Our Myers-Perry-like solution fills this gap, providing the first exact rotating black hole solution with a purely \(r\)-directed bumblebee field (see Eq.~\ref{ansatz-bee}) in higher-dimensional Einstein--Bumblebee gravity. In addition, the relative simplicity of our solution renders it a particularly attractive toy model for exploring quantum gravity effects.}

\subsection{Thermodynamics}
We now investigate the thermodynamics of the black hole derived within the Bumblebee gravity framework. A variety of black hole solutions have been reported in existing studies of this theory, including Schwarzschild-type and RN-type solutions, and their thermodynamic properties have been thoroughly explored using diverse approaches\cite{An:2024fzf,Liu:2024axg}. Notably, multiple formulations of the first law of black hole thermodynamics exist for identical solutions, where the associated entropies differ by a coefficient dependent on the Bumblebee parameter. In this work, we derive two distinct expressions for the first law applicable to the rotating black hole constructed in the preceding section, employing the Wald formalism and Komar integral method respectively.

\subsubsection{Wald formula}
The rotating black hole solution we obtained admits a null Killing vector at its event horizon
\begin{equation}
    \xi=\frac{\partial}{\partial t}+\Omega_{+}\frac{\partial}{\partial \phi} \,.
\end{equation}
By applying the null condition, the $\Omega_{+}$ can be derived
\begin{equation}
    \Omega_{+}=-\omega(r_{0})=\frac{2\nu}{r_{+}\sqrt{r_{+}^4+\nu^2}} \,.
\end{equation}
To derive conserved charges and establish the first law of thermodynamics using the Wald formula, we perform metric variations of our black hole solution over the parameter space. By means of the Wald formalism presented in the preceding section, evaluation of the symplectic form at asymptotic infinity leads to
\begin{equation}
    \frac{1}{16\pi}\int k[g,h]\big|_{r\rightarrow\infty}=\sqrt{1+l}[\delta(\frac{3\pi\mu}{4})-\Omega_{+}\delta(\frac{\sqrt{2\mu}\nu\pi}{4})]=\delta M-\Omega_{+}\delta J \,.
\end{equation}
The mass and angular momentum can be directly extracted
\begin{equation}
    M=\frac{3\pi\sqrt{1+l}\mu}{4}~,~~J=\frac{\sqrt{2(1+l)\mu}\nu\pi}{4}.
\end{equation}
Subsequently, we examine the symplectic form on the horizon, leading to
\begin{equation}\label{Hr0Kerr}
\begin{split}
    \frac{1}{16\pi}\int k[g,h]|_{r=r_{+}}&=\frac{\sqrt{1+l}\pi(r_{+}^4-\nu^2)[\nu r_{+}\delta\nu+(\nu^2+3r_{+}^4)\delta r_{+}]}{4r_{+}^3(\nu^2+r_{+}^4)}
   \\&=\sqrt{1+l}\frac{r_{+}^4-\nu^2}{2\pi r_{+}^3\sqrt{(\nu^2+r_{+}^4)}}\delta(\frac{\pi^2r_{+}\sqrt{\nu^2+r_{+}^4}}{2})
\end{split}
\end{equation}
In general, the symplectic form evaluated at the horizon yields the term $T\delta S$, a combination of temperature and entropy appearing in the first law of thermodynamics. And the Hawking temperature can be computed via standard methods
\begin{equation}\label{T1}
    T=\frac{\kappa}{2\pi}=\frac{r_{+}^4-\nu^2}{2\pi r_{+}^3\sqrt{(\nu^2+r_{+}^4)(1+l)}}\,,
\end{equation}
where $ 	\kappa=\sqrt{-\frac{1}{2}\nabla^{\mu}\xi^{\nu}\nabla_{\mu}\xi_{\nu}} $ is the surface gravity. 
Once the temperature is known, the black hole entropy can be extracted via evaluation of the horizon symplectic form, as follows
\begin{equation}
    S=\frac{\pi^2r_{+}\sqrt{\nu^2+r_{+}^4}}{2}(1+l)\,.
\end{equation}
They comply with the first law of black hole thermodynamics
\begin{equation}
    \delta M=T\delta S+\Omega_{+}\delta J\,.
\end{equation}
The corresponding Smarr relation also holds
\begin{equation}
    M=\frac{3}{2}(TS+\Omega J)\,.
\end{equation}

In contrast to the Einstein framework, the Lorentz-violating parameter 
l
introduces modifications to all thermodynamic quantities in the form of a global scaling factor. Such corrections are vital for investigating the microscopic entropy in later parts of this work. When taking the limit $l\rightarrow 0$, the present results revert to the well-known results of Einstein gravity \cite{Feng:2016dbw}.

It should be noted that multiple distinct formulations of black hole thermodynamics exist within Bumblebee gravity. For example, the authors in Ref.\cite{Liu:2024axg} directly interpret the parameter $m$ as the physical mass of the black hole, which leads to different expressions for black hole entropy. Motivated by this, we explore the microscopic origin of black hole entropy from a holographic perspective via analyzing asymptotic symmetries. Furthermore, our five-dimensional rotating black hole solution provides a nontrivial model for the Kerr/CFT correspondence.

\subsubsection{Komar integration}

In the previous subsection, the thermodynamic entropy is derived from the covariant phase space formalism. However, in theories with non-minimal couplings, the identification of conserved charges is not unique. Komar integrals offer a covariant, coordinate-independent approach to calculating conserved quantities in spacetimes with Killing symmetries. They are fundamental to defining mass and angular momentum in general relativity, and play a vital role in black hole thermodynamics and various modified gravity theories.

The Komar integral approach entails several subtleties, as our black hole is no longer Ricci-flat. This means the associated 2-form $-\star dk$ is not closed. Even so, we may still interpret its integral at infinity as the total energy of the spacetime, by analogy with the Reissner–Nordström black hole. Since the system is invariant under time translations, it possesses a timelike Killing vector $k_{1}=\partial_{t}$. Using the Myers-Perry-like metric (\ref{ansatz}), we can directly derive the corresponding Komar 2-form, namely

\begin{equation}
	-\star dk_{1}=-\frac{r^3\sqrt{f(r)}(-4W(r)h'(r)+4h(r)W'(r)+r^2W(r)^3\omega(r)\omega'(r))}{32\sqrt{h(r)}W(r)}\sin\theta d\theta\wedge d\phi\wedge d\psi+...
\end{equation}
where $...$ stands for terms involving $dr$. Since we integrate the $2$-form on a constant $r$ surface, such terms play no role. The Komar mass is then readily derived as follows:
\begin{equation}
	\hat{M}=\frac{(D-2)}{16\pi (D-3)}\int-\star dk_1=\frac{3\pi \mu}{4\sqrt{1+l}}.
\end{equation}
For the symmetry along the polar direction $k_{2}=\partial_{\psi}$, we have
\begin{equation}
	-\star dk_{2}=\frac{r^2\sqrt{f(r)}W(r)^2\omega'(r)}{32\sqrt{h(r)}}\sin\theta d\theta\wedge d\phi\wedge d\psi \,.
\end{equation}
It corresponds to the angular momentum of the spacetime, namely
\begin{equation}
	\Hat{J}=\frac{1}{16\pi}\int -\star dk_{2}=\frac{\sqrt{2\mu}\nu\pi}{4\sqrt{1+l}}.
\end{equation}
The temperature and angular velocity are derived using standard methods 
\begin{equation}
	\Omega_{+}=\frac{2\nu}{r_{+}\sqrt{r_{+}^4+\nu^2}}~,~~T=\frac{r_{+}^4-\nu^2}{2\pi r_{+}^3\sqrt{(\nu^2+r_{+}^4)(1+l)}}.
\end{equation}
The entropy of the black hole is given by the Bekenstein–Hawking formula
\begin{equation}
	\Hat{S}=\frac{\pi^2r_{+}\sqrt{\nu^2+r_{+}^4}}{2}.
\end{equation}
Within this thermodynamic framework, all relevant quantities are computed independently and satisfy the first law of thermodynamics
\begin{equation}
	\delta \hat{M}=T\delta \hat{S}+\Omega_{+}\delta \hat{J}.
\end{equation}
The associated Smarr formula is also satisfied
	\be
	\hat{M}=\frac{3}{2}(T\hat{S}+\Omega_+ \hat{J})\,.
	\ee
\subsection{Rotating Black Hole in Bumblebee gravity with Cosmological Constant}
Having investigated pure bumblebee gravity, we proceed to incorporate a cosmological constant $\Lambda$ and examine the extended theory accordingly.  In this extended Bumblebee–$\Lambda$ theory, the cosmological constant couples to the symmetry-breaking mechanism, modifying the effective potential of the bumblebee field and influencing black hole solutions, their thermodynamic properties, and the realization of Lorentz violation in a curved background with nontrivial asymptotic geometry.

\subsubsection{Blackhole solution}
Explicitly, we introduce a negative cosmological constant $\Lambda=-6/L^2$ into the theory, with $L$ representing the AdS radius. The equations of motion remain in accordance with our earlier definition (\ref{EOM-gr}). To obtain physically consistent solutions, we introduce a Lagrange multiplier potential, 
\begin{equation}
	V(B^{\mu}B_{\mu}\pm b^2)=\frac{\lambda}{2} (B^{\mu}B_{\mu}\pm b^2),
\end{equation}
where $\lambda$ is a Lagrange multiplier. As a consequence, the vacuum conditions are modified as
\begin{equation}
	\begin{split}
		&V(B_{\mu}B^{\mu}\pm b^2)\Big|_{B^{\mu}=b^{\mu}}=0,
		\\&V'(B_{\mu}B^{\mu}\pm b^2)\Big|_{B^{\mu}=b^{\mu}}=\frac{\lambda}{2} \,.
	\end{split}
\end{equation}
We show that this theory supports rotating AdS-like black holes described by the metric ansatz (\ref{ansatz}). Notably, the rotating solution shares the same form as its asymptotically flat analog. In particular, the metric functions satisfy
\begin{equation}\label{fr}
	f(r) = \frac{h(r)}{1+l} \,.
\end{equation}
From the bumblebee field equations $E_{\nu}=0$, we arrive at a constraint for the Lagrange multiplier $\lambda$, namely
\begin{equation}\label{lambda}
	\lambda=\frac{2\gamma\Lambda}{3\kappa(1+l)}.
\end{equation}
With $\lambda$ determined, the other metric functions $h(r),W(r),\omega(r)$ are straightforward to compute. Their forms coincide with those in Einstein gravity coupled to a cosmological constant
\begin{equation}
	h(r)=W(r)(1+\frac{r^2}{L^2})-\frac{2\mu}{r^2}~,~~W(r)=1+\frac{\nu^2}{r^4}~,~~\omega(r)=-\frac{2\sqrt{2\mu}\nu}{r^4+\nu^2}.
\end{equation}

It is well known that rotating solutions in Einstein gravity take a remarkably simple and uniform form in odd-dimensional spacetimes when all angular momenta are equal. This motivates us to conjecture that analogous rotating black hole solutions with equal angular momenta should also exist in odd-dimensional bumblebee gravity with a cosmological constant. Our verification confirms this expectation, and the detailed results are presented in the appendix A.

\subsubsection{Thermodynamics }
We now proceed to explore the thermodynamics of the AdS black hole. The Hawking temperature and angular velocity are derived using standard methods, which read
\bea
	T&=&\frac{\kappa}{2\pi}=\frac{\sqrt{f'(r_{0})h'(r_0)}}{2\pi\sqrt{W(r_{0})}}=\frac{L^2(r_+^4-\nu^2)+2r_+^6}{2\pi L^2r_+^3\sqrt{(1+l)(r_+^4+\nu^2})^2} \,, \cr
	\Omega_{+}&=&\frac{2\nu}{r_{+}}\sqrt{(1+\frac{r_+^2}{L^2})\frac{1}{r_{+}^4+\nu^2}} \,.
\eea

First, we employ the CPSM to study the conserved charges of this black hole. It is known that the cosmological constant does not contribute directly to the symplectic 2-form $k_{\xi}[g,h]$. Evaluating this quantity at infinity yields
\begin{equation}
	\begin{split}
		\frac{1}{16\pi}\int k[g,h]\big|_{r\rightarrow\infty}&=\sqrt{1+l}[\delta(\frac{\pi(6\mu L^2+\nu^2)}{8L^2})-\Omega_{+}\delta(\frac{\pi\sqrt{2\mu}\nu}{4})]
		\\&=\delta M-\Omega_{+}\delta J,
	\end{split}
\end{equation}
with
\begin{equation}
	M=\frac{\pi\sqrt{1+l}(6\mu L^2+\nu^2)}{8L^2}~,~~J=\frac{\pi\sqrt{1+l}\sqrt{2\mu}\nu}{4}.
\end{equation}
Then evaluating $k_{\xi}[g,h]$ on the horizon gives the structure $T\delta S$
\begin{equation}
	\begin{split}
		\frac{1}{16\pi}\int k[g,h]\big|_{r\rightarrow r_{+}}&=\sqrt{1+l}\big[\frac{L^2(r_+^4-\nu^2)+2r_+^6}{2\pi L^2r_+^3\sqrt{r_+^4+\nu^2}}\delta(\frac{\pi^2 r_{+}\sqrt{r_{+}^4+\nu^2}}{2})\big]
		\\&=T\delta S.
	\end{split}
\end{equation}
Since the temperature $T$ is known, the black hole entropy can be read off directly
\begin{equation}
	S=\frac{\pi^2 r_{+}\sqrt{r_{+}^4+\nu^2}}{2}(1+l) \,.
\end{equation}
All thermodynamic quantities are thus obtained, and they automatically satisfy the first law of thermodynamics
\begin{equation}
	\delta M=T\delta S +\Omega\delta J \,.
\end{equation}

When taking the limit \(l\rightarrow 0\), our expressions reduce to the known results for Einstein gravity with a cosmological constant \cite{Feng:2016dbw}. In the flat limit \(L\rightarrow\infty\), we recover the results derived earlier.

We have calculated the thermodynamic quantities of rotating black holes via the CPSM and the Komar integral in the previous section. Unfortunately, the Komar integral is not applicable in the presence of a negative cosmological constant, as it leads to divergent quantities. Given that the thermodynamic charges evaluated by the Wald formula and the Komar integral differ by an overall factor, we assume this relation persists for Bumblebee gravity theory with a negative cosmological constants and establish an alternative thermodynamic framework
\bea
\hat{M}=\frac{\pi(6\mu L^2+\nu^2)}{8 \sqrt{1+l} L^2}~\,,~~\hat{J}=\frac{\pi\sqrt{2\mu}\nu}{4 \sqrt{1+l}}\,,\quad \hat{S}=\frac{\pi^2 r_{+}\sqrt{r_{+}^4+\nu^2}}{2} \,.
\eea
As usual, standard methods yield the Hawking temperature and angular velocity
\bea
	T=\frac{L^2(r_+^4-\nu^2)+2r_+^6}{2\pi L^2r_+^3\sqrt{(1+l)(r_+^4+\nu^2})^2} \,, \quad \Omega_{+}=\frac{2\nu}{r_{+}}\sqrt{(1+\frac{r_+^2}{L^2})\frac{1}{r_{+}^4+\nu^2}} \,.
\eea
One can readily verify that these thermodynamic quantities obey the first law of black hole thermodynamics.

\section{Kerr/CFT in Bumblebee gravity}
The Kerr/CFT correspondence is a holographic duality inspired by the AdS/CFT correspondence. It states that the near-horizon physics of extremal Kerr black holes is dual to a two-dimensional conformal field theory. The black hole entropy can be exactly reproduced via the Cardy formula of the dual CFT. This framework also extends to non-extremal Kerr black holes and various generalized rotating black holes in modified gravity theories\cite{Castro:2010fd,Krishnan:2009tj,Liu:2021hvb}. We have successfully obtained rotating black hole solutions within Bumblebee gravity in the preceding part. Here we study the associated Kerr/CFT correspondence of these solutions.

\subsection{A Brief Review of Kerr/CFT for Kerr black hole in Einstein gravity}
Before proceeding to study the microscopic entropy of our solutions, we present a short review of the Kerr/CFT correspondence, which forms the theoretical basis for the following analysis.

According to the Kerr/CFT correspondence, the geometry in the vicinity of the horizon of an extremal rotating black hole is dual to a two-dimensional conformal field theory. This duality enables the Bekenstein–Hawking entropy to be recovered via the Cardy formula. The line element of the four-dimensional Ricci-flat Kerr black hole, written in Boyer–Lindquist coordinates, is expressed as\cite{Kerr:1963ud}
\begin{equation}
    ds^2=-\frac{\Delta}{\rho^2}(d\hat{t}-a\sin^2\theta d\hat{\phi})^2+\frac{\sin^2\theta}{\rho^2}((\hat{r}^2+a^2)d\hat{\phi}-ad\hat{t})^2+\frac{\rho^2}{\Delta}d\hat{r}^2+\rho^2d\theta^2.
\end{equation}
where
\begin{equation}
    \rho^2=\hat{r}^2+a^2\cos^2\theta~,~~\Delta=\hat{r}^2-2m\hat{r}+a^2.
\end{equation}
The metric has two integration constants: the mass parameter $m$ and the rotation parameter $a$. The outer horizon is situated at \(\hat{r}=r_{+}\), the largest root of \(\Delta=0\). We present the Hawking temperature, horizon angular velocity, Bekenstein–Hawking entropy, angular momentum and mass as follows
\begin{equation}
    T=\frac{r_{+}^2-a^2}{4\pi r_{+}(r_{+}^2+a^2)}~,~~\Omega_{+}=\frac{a}{r_{+}^2+a^2}~,~~S=\pi(r_{+}^2+a^2)~,~~J=ma~,~~M=m.
\end{equation}
They satisfy the first law of thermodynamics
\begin{equation}
    \delta M=T\delta S+\Omega_{+}\delta J.
\end{equation}
In the extremal limit \(T=0\), we have \(r_{+}=M=\sqrt{J}\) for these parameters. The metric for the near-horizon extremal Kerr (NHEK) geometry takes the form
\begin{equation}
    ds^2=2J\Omega^2(-(1+r^2)d\tau^2+\frac{dr^2}{1+r^2}+d\theta^2+\Lambda^2(d\phi+rd\tau)^2),
\end{equation}
where
\begin{equation}
    \Omega^2=\frac{1+\cos^2\theta}{2}~,~~\Lambda=\frac{2\sin\theta}{1+\cos^2\theta}.
\end{equation}
Each fixed-\(\theta\) slice corresponds to a quotient of warped \(\text{AdS}_3\), with the isometry group \(\text{SL}(2,\mathbb{R})\times \text{U}(1)\) \cite{Bardeen:1999px}. As presented in \cite{Guica:2008mu}, proper boundary conditions at \(r=\infty\) must be specified to extract the asymptotic symmetry group (ASG) and the associated generators of the near-horizon metric, as given below
\begin{equation}
	\begin{pmatrix}
		h_{\tau\tau}=\mathcal{O}    (r^2)&h_{\tau\phi}=\mathcal{O}(1)&h_{r\theta}=\mathcal{O}(\frac{1}{r})&h_{\tau r}=\mathcal{O}(\frac{1}{r^2})\\...&h_{\phi\phi}=\mathcal{O}(1)&h_{\phi\theta}=\mathcal{O}(\frac{1}{r})&h_{\phi r}=\mathcal{O}(\frac{1}{r})\\...&...&h_{\theta\theta}=\mathcal{O}(\frac{1}{r})&h_{\theta\phi}=\mathcal{O}(\frac{1}{r^2})\\...&...&...&h_{rr}=\mathcal{O}(\frac{1}{r^3})
	\end{pmatrix}.
\end{equation}
Here \(h_{\mu\nu}\) represents the perturbation around the background metric \(\bar{g}_{\mu\nu}\), such that \(g_{\mu\nu}=\bar{g}_{\mu\nu}+h_{\mu\nu}\). Under these boundary conditions, we can directly calculate the generators of the ASG
\begin{equation}
    \xi_{\epsilon}=\epsilon(\phi)\frac{\partial}{\partial\phi}-r\epsilon'(\phi)\frac{\partial}{\partial r}.
\end{equation}
Since \(\epsilon(\phi)\) is a periodic function of the angular coordinate \(\phi\), we expand it into a Fourier mode \(\epsilon=-e^{-in\phi}\). The diffeomorphisms take the form
\begin{equation}
    \xi_{n}=-e^{-in\phi}\frac{\partial}{\partial\phi}-inre^{-in\phi}\frac{\partial}{\partial r}.
\end{equation}
The commutator of the corresponding Killing vector fields forms the Witt algebra
\begin{equation}
    i[\xi_{m},\xi_{n}]=(m-n)\xi_{m+n}.
\end{equation}
As stated in \cite{Guica:2008mu}, the conserved charges for the diffeomorphisms are obtained by integrating the 2-form \(k_{\xi}\) at spatial infinity \(r\rightarrow\infty\)
\begin{equation}
    \delta Q_{\xi}=\frac{1}{16\pi}\int_{\partial\mathcal{M}}k_{\xi}.
\end{equation}
The 2-form for pure Einstein gravity can be derived from the covariant phase space formalism in Eq.(\ref{k-bee}) by setting \(B_{\mu}=0\), namely
\begin{equation}
     k_{\xi}=\epsilon_{\alpha\beta\mu\nu}(-\frac{1}{2}h\nabla^{\nu}\xi^{\mu}-h^{\sigma\mu}\nabla_{\sigma}\xi^{\nu}+\xi_{\sigma}\nabla^{\mu}h^{\nu\sigma}+\xi^{\mu}\nabla^{\nu}h-\xi^{\mu}\nabla_{\sigma}h^{\nu\sigma})dx^\alpha\wedge dx^{\beta}.  
\end{equation}
The Dirac bracket for the charges generates the central extension of the algebra
\begin{equation}
    \{Q_{\xi_{m}},Q_{\xi_{n}}\}=Q_{[\xi_{m},\xi_{n}]}+\frac{1}{16\pi}\int_{\partial\mathcal{M}}k_{\xi_{m}}[\mathcal{L}_{\xi_{n}}g,g] \,,
\end{equation}
where \(\mathcal{L}_{\xi_{n}}\) stands for the Lie derivative acting on the background metric \(g_{\mu\nu}\), satisfying \(\mathcal{L}_{\xi_{n}}g_{\mu\nu}=\nabla_{\mu}\xi_{(n)\nu}+\nabla_{\nu}\xi_{(n)\mu}\). The central charge is extracted from the second term. Inserting the NHEK metric into the central term yields the result below
\begin{equation}
    \frac{1}{16\pi}\int_{\partial\mathcal{M}}k_{\xi_{m}}[\mathcal{L}_{\xi_{n}}g,g]=-i(m^3+2m)J\delta_{m+n,0}.
\end{equation}
The resulting quantum algebra is the Virasoro algebra
\begin{equation}
    [L_{m},L_{n}]=(m-n)L_{m+n}+\frac{c}{12}(m^3+\alpha m)\delta_{m+n,0},
\end{equation}
where the linear term in $m$ is irrelevant, because such contributions may be removed by c-number shifts of the generators
\begin{equation}
    \Tilde{L}_{m}=L_{m}+Nc\delta_{m}~,~~N\in \mathbb{R}.
\end{equation}
Thus the constant \(\alpha\) is irrelevant here. The central charge is directly given by
\begin{equation}
c=12J,
\end{equation}
which is proportional to the angular momentum. To compute the microscopic entropy, we adopt the Frolov-Thorne vacuum \cite{Frolov:1989jh} to define the vacuum state for the extremal Kerr black hole. The dimensionless Frolov-Thorne temperature for the NHEK geometry reads
\begin{equation}
T_{FT}=\frac{1}{2\pi}.
\end{equation}A key result of the Kerr/CFT correspondence is that the microscopic entropy calculated via the Cardy formula
\begin{equation}
S_{C}=\frac{\pi^2}{3}cT_{FT}=2\pi J,
\end{equation}
precisely agrees with the Bekenstein–Hawking entropy of the extremal Kerr black hole. This excellent match indicates that the fundamental degrees of freedom underlying black hole entropy are fully described by the dual CFT.

In subsequent sections, we implement this formalism within Lorentz-violating gravity and check if the duality remains valid.

\subsection{Kerr/CFT Correspondence for rotating black holes in Bumblebee gravity}
Though the Kerr/CFT correspondence for BTZ-like black holes in Bumblebee gravity has been investigated in the literature\cite{Ding:2025cno}, this duality is physically trivial. It provides no novel insights into graviton dynamics in three-dimensional spacetime, which accounts for its limited research significance. Furthermore, we find that the Kerr/CFT-derived entropy of BTZ-like black holes reported in existing literature is unreliable, and we present a comprehensive discussion of this issue in the Appendix B. In this section, we conduct a thorough analysis of the Kerr/CFT correspondence for the newly derived five-dimensional rotating black hole in Bumblebee gravity.

\subsubsection{Near Horizon Geometry}
To explore the Kerr/CFT correspondence, we construct the near-horizon geometry of the extremal black hole. For an extremal black hole, the inner and outer horizons \(r_{\pm}\) coincide at a single degenerate horizon \(r_{0}\). As a result, the black hole temperature vanishes. In this case, the relation among the horizon radius, mass parameter and rotation parameter satisfies the following equality:
\begin{equation}
    r_{0}=\sqrt{\mu}=\sqrt{\nu}.
\end{equation}

Near the horizon of the extremal black hole, the functions \(h,f,W,\omega\) take the form
\begin{equation}
\begin{split}
    &W=2~,~~\omega(r)=\omega(r_{0})+\omega'(r_{0})(r-r_0)~,\\&f=\frac{h}{1+l}~,~~
    h=V(r-r_{0})^2+\mathcal{O}(r-r_{0})^3,
\end{split}
\end{equation}
where
\begin{equation}
    V=\frac{h''(r_{0})}{2!}=\frac{4}{\nu}~,~~\omega'(r_{0})=\frac{2\sqrt{2}}{\nu}
\end{equation}
To extract the near-horizon geometry, we first perform the coordinate transformations
\begin{equation}
    r=r_0(1+\lambda y)~,~~ \psi=\tilde{\psi}-\alpha t~,~~\alpha=\omega(r_{0}).
\end{equation}
The metric becomes
\begin{equation}
    ds^2=-\frac{Vr_{0}^2\lambda^2}{2}y^2dt^2+\frac{1+l}{Vy^2}dy^2+\frac{\nu}{2}(d\tilde{\psi}+\cos\theta d\phi+\omega'(r_{0})r_{0}\lambda ydt)^2+\frac{r^2}{4}(d\theta^2+\sin^2\theta d\phi^2).
\end{equation}
After rescaling the time coordinate via
\begin{equation}
	t\rightarrow \tilde{t}=\frac{V\lambda r_{0}}{\sqrt{2(1+l)}}t,
\end{equation}
we obtain the near-horizon metric
\begin{equation}
    ds^2=\frac{1}{V}[-(1+l)y^2d\tilde{t}^2+\frac{1+l}{y^2}dy^2+d\theta^2+\sin^2\theta d\phi^2+2(d\tilde{\psi}+\cos\theta d\phi+\sqrt{1+l}yd\tilde{t})^2].
\end{equation}
These are Poincaré-type coordinates that cover only a portion of the AdS geometry. We therefore transform the AdS\(_2\) Poincaré coordinates \((y,t)\) to the global AdS coordinates \((r',\tau)\). Following the discussion in \cite{Lu:2008jk}, we adopt the coordinate transformation
\begin{equation}
    y=r'+\sqrt{1+r'^2}\cos\tau~,~~t=\frac{\sqrt{1+r^2}\sin\tau}{r+\sqrt{1+r^2}\cos\tau},
\end{equation}
the metric becomes
\begin{equation}
    ds^2=\frac{1}{V}[-(1+l)(1+r'^2)d\tau^2+\frac{(1+l)dr'^2}{1+r'^2}+d\theta^2+\sin^2\theta d\phi^2+2(d\tilde{\psi}+\cos\theta d\phi+\sqrt{1+l}r'd\tau+d\gamma)^2],
\end{equation}
where
\begin{equation}
   d\gamma= d(\sqrt{(1+l)}\log(\frac{1+\sqrt{1+r^2}\sin\tau}{\cos\tau+r\sin\tau})).
\end{equation}
The term $d\gamma$ can be eliminated by the redefinition of the angular coordinate 
\begin{equation}
    d\tilde{\psi}=d\Bar{\psi}-d\gamma.
\end{equation}
And then re-scaling the time coordinate $\tau\rightarrow\tau/\sqrt{1+l}$, the metric of the near-horizon geometry finally takes the form of global AdS
\begin{equation}
    ds^2=\frac{1}{V}[-(1+r'^2)d\tau^2+\frac{(1+l)dr'^2}{1+r'^2}+d\theta^2+\sin^2\theta d\phi^2+2(d\Bar{\psi}+\cos\theta d\phi+r'd\tau)^2].
\end{equation}
The metric is homogeneous and describes a constant \(\text{U}(1)\) bundle over \(\text{AdS}_2\times\text{S}^2\), which serves as the standard setup for the Kerr/CFT correspondence. Note that the Lorentz-violating parameter only rescales the radial component and leaves the global structure of the near-horizon geometry intact. Finally, under the above coordinate transformations, the bumblebee field takes the form
\begin{equation}
\begin{split}
      B&=\sqrt{b^2(1+l)}\frac{dr}{\sqrt{V(r-r_{0})^2}}=\sqrt{\frac{b^2(1+l)}{V}}\frac{dy}{y}
      \\&=\sqrt{\frac{b^2(1+l)}{V}}[\frac{1+\frac{r'\cos\tau}{\sqrt{1+r'^2}}}{r'+\sqrt{1+r'^2}\cos\tau}dr'-\frac{\sqrt{1+r'^2}\sin\tau}{r+\sqrt{1+r'^2}\cos\tau}d\tau].  
\end{split}
\end{equation}

\subsubsection{Central Charge and Cardy Formula}
We proceed to analyse the asymptotic symmetry group (ASG) of the near-horizon geometry. Adopting the method outlined in \cite{Guica:2008mu}, we investigate the ASG for this five-dimensional extremal black hole. The boundary conditions coincide with those presented earlier \cite{Lu:2008jk}. The Lorentz-violating factor $l$ merely induces a constant shift of the \(h_{rr}\) component. Similar to the AdS\(_3\) scenario, such a shift has no effect on the ASG-generated diffeomorphisms. The corresponding diffeomorphisms are given by
\begin{equation}\label{killing-cft}
    \xi_{m}=-e^{-im\Bar{\psi}}\frac{\partial}{\partial\Bar{\psi}}-imr'e^{-im\Bar{\psi}}\frac{\partial}{\partial r'},
\end{equation}
where the Witt algebra automatically satisfies 
\begin{equation}
    i[\xi_{m},\xi_{n}]=(m-n)\xi_{m+n}.
\end{equation}
We adopt the Lie derivative\(\mathcal{L}_{\xi_m}g_{\mu\nu}=\nabla_{\mu}\xi_{(m)\nu}+\nabla_{\nu}\xi_{(m)\mu}\)acting on the background metric, and apply the covariant phase space formalism to compute the conserved charges in the limit \(r'\to\infty\). This yields the central extension of the Virasoro algebra
\begin{equation}
    \frac{1}{16\pi}\int k_{\xi_{n}}[g,h]=\frac{-i}{16\pi}(\frac{\nu^{3/2}\sqrt{1+l}}{2\sqrt{2}})(2m+m^3)\times 16\pi^2\delta_{m+n,0}.
\end{equation}
We focus on the term appropriately to $m^3$ 
\begin{equation}
   \frac{i}{16\pi}\int k_{\xi_{n}}[g,h]\Big|_{m^{3}}=\frac{\pi m^3\nu^{3/2}\sqrt{1+l}}{2\sqrt{2}}=\frac{cm^3}{12},
\end{equation}
which yields the central charge
\begin{equation}
    c=\frac{6\pi\nu^{3/2}\sqrt{1+l}}{\sqrt{2}}.
\end{equation}
The central charge is modified by the Lorentz-violating factor \(\sqrt{1+l}\), a behaviour consistent with pure AdS\(_3\). We employ the Frolov-Thorne vacuum\cite{Frolov:1989jh} to construct a well-defined vacuum for the near-horizon extremal rotating black hole and proceed to evaluate the Cardy entropy. At extremality, the left- and right-moving temperatures are given by\cite{Liu:2021hvb}:
\begin{equation}
    T_{L}=\frac{V}{2\pi\omega'(r_{0})\sqrt{(1+l)W_{0}}}=\frac{1}{2\pi\sqrt{1+l}}~,~~T_{R}=0.
\end{equation}
Substituting this result into the Cardy formula, we arrive at the corresponding entropy: microscopic entropy
\begin{equation}
    S_{C}=\frac{\pi^2\nu^{3/2}}{\sqrt{2}}.
\end{equation}
This shows that the entropy satisfies \(S_{C}=A/4\), one quarter of the degenerate event horizon area, which matches the Bekenstein–Hawking entropy.
The entropy derived from the Kerr/CFT result supports the second formulation of the first law of black hole thermodynamics, where the conserved quantities are evaluated via Komar integration,
\begin{equation}
    \hat{S}=\frac{A}{4}=S_{C}.
\end{equation}
Nevertheless, the Kerr/CFT entropy differs from the thermodynamic entropy from the Wald formula by a factor of \(1+l\), namely
\begin{equation}
    S=(1+l)S_{C}.
\end{equation}

Since both the asymptotic symmetry group and the near-horizon geometry remain intact, this discrepancy does not stem from changes in the dual CFT. It instead implies that the identification of thermodynamic quantities in Lorentz-violating gravity calls for further careful examination.

\subsection{Kerr/CFT for Bumblebee gravity with cosmological constant}
We now generalize the analysis to spacetimes with a negative cosmological constant. This provides a nontrivial consistency test of our prior findings and enables us to explore whether the entropy mismatch remains present under more general conditions.

The rotating black hole solution in Bumblebee gravity with a cosmological constant was derived in the previous section. We now proceed to analyze the Kerr/CFT correspondence for this black hole in the extremal limit. In the extremal case \(r_{\pm}=r_{0}\), the relevant parameters satisfy
\begin{equation}
    \nu=\frac{r_{0}^2\sqrt{L^2+2r_{0}^2}}{L}~,~~\mu=\frac{r_{0}^2(L^2+r_{0}^2)^2}{L^4}.
\end{equation}
Using these parameters, the near-horizon metric functions take the form:
 \begin{equation}
     h=V(r-r_{0})^2+\mathcal{O}((r-r_{0})^3)~,~~f=\frac{h}{1+l}~,~~W=W_{0}~,~~\omega=\omega_{0}+\omega_{1}(r-r_{0}),
 \end{equation}
where
 \begin{equation}
    V=4(\frac{3}{L^2}+\frac{1}{r_{0}^2})~,~~W_{0}=2+\frac{2r_{0}^2}{L^2}~,~~\omega_{1}=\frac{2\sqrt{2}L\sqrt{L^2+2r_{0}^2}}{r_{0}^2(L^2+r_{0}^2)}.
\end{equation}
Following the same procedure adopted in Sec 5.2, the near horizon geometry of metric and bumblebee field are given by
\begin{equation}
    ds^2=\frac{1}{V}[-(1+r'^2)d\tau^2+\frac{(1+l)dr'^2}{1+r'^2}]+\frac{r_{0}^2}{4}(d\theta^2+\sin^2\theta d\phi^2)+\frac{r_{0}^2W_{0}}{4}(d\Bar{\psi}+\cos\theta d\phi+\frac{\sqrt{W_{0}}\omega_{1}r}{V}d\tau)^2,
\end{equation}
and
\begin{equation}
\begin{split}
      B=\sqrt{\frac{b^2(1+l)}{V}}[\frac{1+\frac{r'\cos\tau}{\sqrt{1+r'^2}}}{r'+\sqrt{1+r'^2}\cos\tau}dr'-\frac{\sqrt{1+r'^2}\sin\tau}{r+\sqrt{1+r'^2}\cos\tau}d\tau].
\end{split}
\end{equation}
The near-horizon geometry again presents a warped AdS\(_3\) structure, and the asymptotic symmetry group is governed by a Virasoro algebra identical to those in previous sections. In particular, the Lorentz-violating parameter leaves the structure of the symmetry algebra unchanged. Accordingly, the diffeomorphisms \(\xi_{m}\) remain intact. We extract the term in \(k_{\xi}[g,h]\) proportional to \(m^3\) and read off the central charge
\begin{equation}
    c=\frac{3\sqrt{2}L\pi r_{0}^3\sqrt{L^2+2r_{0}^2}}{L^2+3r_{0}^2}\sqrt{1+l}.
\end{equation}
We find the central charge satisfies \(c\sim\sqrt{1+l}\). Meanwhile, the Frolov–Thorne temperature is given by
\begin{equation}
    T_{L}=\frac{(L^2+3r_{0}^2)\sqrt{L^2+r_{0}^2}}{2L^2\pi\sqrt{(\sqrt{1+l})(L^2+2r_{0}^2)}}~,~~T_{R}=0.
\end{equation}
When the cosmological constant vanishes (\(L\to\infty\)), we recover \(T_{L}=1/(2\pi)\). The Cardy formula then yields the microscopic entropy
\begin{equation}
    S_{CFT}=\frac{\pi^2}{3}cT_{L}=\pi^2 r_{0}^3\sqrt{(\frac{1}{2}+\frac{r_{0}^2}{2L^2})},
\end{equation}

This shows that, up to standard geometric factors, the Cardy entropy satisfies \(S_{C}\sim(1+l)^0\) and equals one quarter of the event horizon area. It reproduces the known black hole entropy and supports the second formulation of the first law of black hole thermodynamics. Meanwhile, it differs from the thermodynamic entropy \(S\sim(1+l)\) derived via the Wald formalism, so we once again find an entropy mismatch. Including a cosmological constant does not alter this behaviour. Instead, it demonstrates that the mismatch is a robust, universal feature of rotating black holes in Einstein–bumblebee gravity. This strongly suggests that the discrepancy is intrinsic to Lorentz-violating dynamics, rather than an artifact of specific solutions.

\section{Conclusion and Discussion}
In this work, we construct rotating black hole solutions in five-dimensional Bumblebee gravity with and without a cosmological constant for the first time. A prominent feature of these solutions is that the Lorentz-violating parameter enters the metric in a remarkably simple form, which effectively rescales the radial component of Myers–Perry black holes while keeping their angular structure intact. Despite this seemingly minor modification, the resulting spacetimes are physically distinct from Myers–Perry black holes in Einstein gravity, as manifested by their nonvanishing curvature invariants.
 
We further investigate the black hole thermodynamics of the constructed solutions using both the Wald formalism and the Komar integral method. Within the covariant phase space formalism, we derive the conserved charges and obtain the first set of thermodynamic quantities. In this formulation, both the mass and angular momentum are rescaled by a factor of $\sqrt{1+l}$. In particular, the entropy acquires a multiplicative correction $1+l$ compared with the standard Bekenstein–Hawking entropy,
\begin{equation}
	S=(1+l)\frac{A}{4}.
\end{equation}
These thermodynamic quantities automatically satisfy the first law of thermodynamics and the Smarr relation. Alternatively, we develop a second thermodynamic description based on the Komar integral approach, in which the Komar mass and angular momentum are scaled by $1/\sqrt{1+l}$. Correspondingly, the entropy strictly obeys the conventional Bekenstein–Hawking area law,
\begin{equation}
	\hat{S}=\frac{A}{4}.
\end{equation}
In this second formulation, all thermodynamic quantities can be independently computed and naturally satisfy both the first law and the Smarr relation.

To investigate which entropy possesses a more self‑consistent microscopic origin, we carried out a full analysis of the Kerr/CFT correspondence for our constructed solutions. In particular, the Virasoro algebra and its central extension preserve their standard structure, except for an overall scaling of the central charge by a factor of $\sqrt{1+l}$. This indicates that the dual conformal field theory is not fundamentally altered by Lorentz violation. By choosing the Frolov–Thorne vacuum, we obtained the microscopic entropy via the Cardy formula. We found that the Cardy entropy equals one‑quarter of the event horizon area, which matches the thermal entropy from the Komar version:
 \begin{equation}
     S_{C}=\frac{A}{4}=\hat{S}.
 \end{equation}
 
 This result is rather striking. Nevertheless, a clear discrepancy persists between this microscopic entropy and the entropy derived via the Wald formalism. Notably, this entropy mismatch also plagues black hole solutions within Horndeski gravity. Both cases have one common feature that the matter fields diverge at the horizon. We stress that the present work neither attempts to adjudicate which entropy definition is physically correct nor resolves the underlying origin of this entropy discrepancy.

Furthermore, a recent viewpoint suggests that, in this class of modified gravity theories, the propagation of gravitational waves is modified\cite{Hajian:2020dcq}. Consequently, the surface gravity and temperature experienced by gravitons should also be modified. According to this viewpoint, only when the modified temperature is used can the entropy inferred from the Wald formalism be interpreted as the physical black hole entropy. In the appendix C, we follow this perspective and present the corresponding results. We find that the black hole entropy constructed on the basis of this argument agrees with the microscopic entropy obtained from the Kerr/CFT correspondence. Nevertheless, we cannot claim that this provides the final answer. We hope that the origin of this entropy discrepancy can be understood at a more fundamental level in future work.

\section{Acknowledgement}
 We are grateful to Yan-Chen Bi and Ming-Chi Ma for the useful discussion. We also thanks M.M. Sheikh-Jabbari for the detailed explanation of \cite{Hajian:2020dcq}. This work is supported in part by NSFC (National Natural Science Foundation of China) Grant No. 12575061, Tianjin University Graduate Liberal Arts and Sciences Innovation Award Program (2023) No. B1 2023-005 and Tianjin University Self-Innovation Fund Extreme Basic Research Project Grant No. 2025XJ21-0007.
\section*{Appendix}
\appendix

\section{Rotating Black Hole in General Odd Dimensions}
In this section, we consider rotating black holes with all angular momenta equal in $(2n+1)$-dimensional Einstein--Bumblebee gravity with a cosmological constant. Specifically, we adopt a linear potential for the bumblebee field
\begin{equation}
    V=\frac{\lambda}{2}(B_{\mu}B^{\mu}\pm b^2).
\end{equation}
The equations of motion are given by
\begin{equation}\label{EOM-bee}
	E_{\nu}=\nabla^{\mu}b_{\mu\nu}+\frac{\gamma}{\kappa}b^{\mu}R_{\mu\nu}-2V'b_{\nu} =0,
\end{equation}
and 
\begin{equation}
	\begin{split}
		E_{\mu\nu}=&R_{\mu\nu}-\frac{1}{2}g_{\mu\nu}R+\Lambda g_{\mu\nu}-\kappa T_{\mu\nu}^{(Bee)}=0,
	\end{split}
\end{equation}
where
\begin{equation}
    \begin{split}
        T_{\mu\nu}^{(Bee)}=&-b_{\mu\alpha}b^{\alpha}_{~\nu}-\frac{1}{4}b_{\alpha\beta}b^{\alpha\beta}-g_{\mu\nu}V+b_{\mu}b_{\nu}\lambda+\frac{\gamma}{\kappa}(\frac{1}{2}b^{\alpha}b^{\beta}R_{\alpha\beta}g_{\mu\nu}-b_{\mu}b^{\alpha}R_{\alpha\nu}-b_{\nu}b^{\alpha}R_{\alpha\mu}\\&+\frac{1}{2}\nabla_{\alpha}\nabla_{\mu}(b^{\alpha} b_{\nu})
        +\frac{1}{2}\nabla_{\alpha}\nabla_{\nu}(b^{\alpha} b_{\mu})-\frac{1}{2}\nabla^2(b_{\mu}b_{\nu})-\frac{1}{2}g_{\mu\nu}\nabla_{\alpha}\nabla_{\beta}(b^{\alpha}b^{\beta})).
    \end{split}
\end{equation}
For the case where all angular momenta are equal, the metric ansatz can be written as
\begin{equation}
    ds^2_{2n+1}=-\frac{h}{W}dt^2+\frac{dr^2}{f}+\frac{r^2W}{4}(\sigma+\omega )^2+ds^2_{\mathbb{CP}^{n-1}},
\end{equation}
where $ds^2_{\mathbb{CP}^{n-1}}$ is the metric of an $(n-1)$ dimensional complex projective space, $\omega=\omega(r)dt$ and $\sigma$ is the 1-form connection along the Hopf fiber direction. Following the same procedure, we find that the metric functions of the black hole solution satisfy
\begin{equation}
    f=\frac{h}{1+l}.
\end{equation}
And the constrain of the Lagrange multiplie (\ref{lambda}) is generalized as
\begin{equation}
   \lambda=\frac{2\gamma\Lambda}{(2n-1)\kappa(1+l)}. 
\end{equation}
By submitting $\Lambda=-n(2n-1)/L^2$ into the equations of motion $E_{\mu\nu}=0$, the equations degenerate to the equations in Einstein gravity. Thus the explicit metric function are given by
\begin{equation}
    h=(1+\frac{r^2}{L^2})W-\frac{2\mu}{r^{2(n-1)}}~,~~W=1+\frac{\nu^2}{r^{2n}}~,~~\omega=-\frac{2\sqrt{2\mu}\nu}{r^{2n}W}dt.
\end{equation}
As a concrete illustration of our general solutions, we now present an explicit example in seven dimensions. The corresponding rotating black hole solution is given by
\begin{align}
	ds^2_{7} &= -\frac{h}{W}dt^2+\frac{(1+l)dr^2}{h}+\frac{r^2W}{4}\bigl(d\psi+\cos\theta_{1}\,d\phi_{1}+\cos\theta_{2}\,d\phi_{2}+\omega\bigr)^2 \nonumber\\
	&\quad +\frac{r^2}{4}\bigl(d\theta_{1}^2+\sin^2\theta_{1}\,d\phi_{1}^2+d\theta_{2}^2+\sin^2\theta_{2}\,d\phi_{2}^2\bigr),
\end{align}
where
\begin{equation}
    h=(1+\frac{r^2}{L^2})(1+\frac{\nu^2}{r^6})-\frac{2\mu}{r^4}~,~~W=1+\frac{\nu^2}{r^6}~,~~\omega=-\frac{2\sqrt{2\mu}\nu}{r^6+\nu^2}dt.
\end{equation}
The thermodynamic quantities can be calculated through the similar procedure.

\section{Microscopic Entropy Discrepancy in AdS$_{3}$/CFT$_{2}$ Correspondence}
In this section, we analyze the holographic properties of Lorentz-violating gravity in three dimensions. The BTZ-like black hole provides a particularly simple setup in which the asymptotic symmetry group and the dual CFT can be constructed explicitly. As we will show, this example also reveals a nontrivial discrepancy between the thermodynamic and microscopic entropy.

\subsection{BTZ-like Black Hole}
We consider three-dimensional Einstein--Bumblebee gravity with a linear potential for the bumblebee vector field
\begin{equation}
	V(X)=\frac{\lambda X}{2} \,,
\end{equation}
which enforces the vacuum constraint and $\lambda$ is a Lagrange-multiplier. The rotating BTZ-like solution for this theory was previously obtained in \cite{Ding:2023niy}
\begin{equation}
	\begin{split}
		&ds^2=-h(r)dt^2+\frac{dr^2}{f(r)}+r^2(d\phi-\frac{j}{2r^2}dt)^2~,~~b_{\mu}=(0,b_{r}(r),0),
		\\&    h(r)=-m-\Lambda r^2+\frac{j^2}{4r^2}~,~~f(r)=\frac{h(r)}{1+l}~,~~b_{r}(r)=\sqrt{\frac{b^2}{f(r)}}.
	\end{split}
\end{equation}
where $l$ denotes the Lorentz symmetry breaking parameter, which is defined as $l=\gamma b^2$ and the negative cosmological constant can be written as $\Lambda=-1/L^2$ . The event horizon is located at $r_{0}$ which is determined by $h(r_{0})=0$. The black hole exist a null Killing vector on horizon
\begin{equation}
	\xi=\frac{\partial}{\partial t}+\Omega\frac{\partial}{\partial\phi}.  
\end{equation}
The corresponding temperature and angular velocity at the horizon are then given by
\begin{equation}\label{T-btz}
	\Omega=\frac{j}{2r_{0}^2}~,~~T=\frac{\kappa}{2\pi}=\frac{h'(r_{0})}{4\pi\sqrt{1+l}}=-\frac{j^2+4r_{0}^4\Lambda}{8\pi r_{0}^3\sqrt{1+l}},
\end{equation}
where $\kappa$ is the surface gravity. By using the CPSM, we can calculate the symplectic $2$-form in the infinity
\begin{equation}
	\frac{1}{16\pi}\int k_{\xi}[\delta g,g]\Big|_{r=\infty}=\frac{\sqrt{1+l}}{8}(\delta m-\Omega\delta j)=\delta M-\Omega\delta J.
\end{equation}
We can read-off the mass and angular momentum
\begin{equation}
	M=\frac{\sqrt{1+l}m}{8}~,~~J=\frac{\sqrt{1+l}j}{8}.
\end{equation}
And the syyplectic $2$-form on the horizon gives $T\delta S$
\begin{equation}
	\frac{1}{16\pi}\int k_{\xi}[\delta g,g]\Big|_{r=r_{0}}=-\frac{\sqrt{1+l}(j^2+4r_{0}^2\Lambda)}{16r_{0}^3}\delta r_{0}=T\delta S.
\end{equation}    
We can thus obtain the entropy
\begin{equation}
	S=\frac{r_{0}\pi(1+l)}{2}.
\end{equation}
The first law is automatically satisfied
\begin{equation}
	\delta M=T\delta S+\Omega\delta J.
\end{equation}
\subsection{ASG}
To analyze the dual CFT, we now determine the asymptotic symmetry group (ASG). Expand the metric in Fefferman–Graham coordinates,
\begin{equation}
	ds^2=\frac{L^2(1+l)dr^2}{r^2}+\frac{r^2}{L^2}(g^{(0)}_{ij}dx^idx^j+\mathcal{O}(\frac{1}{r})).
\end{equation}
In this framework, we impose the Brown-Henneaux boundary condition, which the $g^{(0)}_{ij}=\eta_{ij}$ is fixed. Let us consider an arbitrary diffeomorphism $\xi=\xi(r,x^{i})$, and the Lie derivative of metric is given by
\begin{equation}
	\delta_{\xi}g_{\mu\nu}=\xi^{\rho}\partial_{\rho}g_{\mu\nu}+g_{\rho\nu}\partial_{\mu}\xi^{\rho}+g_{\mu\rho}\partial_{\nu}\xi^{\rho}.
\end{equation}
In order to obtain the specific form of the diffeomorphism $\xi$, we first see the $rr$-component
\begin{equation}
	2g_{rr}\partial_{r}\xi^{r}+\xi^{r}\partial_{r}g_{rr}=0.
\end{equation}
We define 
\begin{equation}
	\xi^{r}=\tilde{R}(r,x^{i}),
\end{equation}
it gives
\begin{equation}
	\partial_{r}\tilde{R}-\frac{\tilde{R}}{r}=0~~\rightarrow \tilde{R}=rR(x^i).
\end{equation}
Then the $r,i$ components of Lie derivative are
\begin{equation}
	\begin{split}
		\delta_{\xi}g_{ri}&=g_{ij}\partial_{r}\xi^{j}+rg_{rr}\partial_{i}R(x^i)=0~~\rightarrow ~~\xi^{i}=V^{i}(x^i)-\int rg_{rr}g^{ik}\partial_{k}R.
	\end{split}    
\end{equation}
Then we focus on the $ij$ components
\begin{equation}
	\begin{split}
		\delta_{\xi}g_{ij}=&R(x^i)\frac{2r^2}{L^2}g_{ij}^{(0)}+\frac{r^2}{L^2}(V^k\partial_{k}g_{ij}^{(0)}+g_{ik}^{(0)}\partial_{j}V^{k}+g_{ik}^{(0)}\partial_{j}V^k)\\&-\partial_{k}g_{ij}\int rg_{rr}g^{kl}\partial_{l}R-2g_{k(i}\partial_{j)}\int rg_{rr}g^{kl}\partial_{l}R,
	\end{split}
\end{equation}
Taking the trace of the equation and ignoring the higher-order fall-off terms, one can obtain
\begin{equation}
	R(x^i)=-\frac{\partial_{i}V^i(x^i)}{2}.
\end{equation}
We then use a light-cone coordinate $x_{+}=t/L+\phi,~x_{-}=t/L-\phi$ to rewrite the $g_{ij}^{(0)}$
\begin{equation}
	ds_{(0)}^2=-dt^2+L^2d\phi^2=-L^2dx_{+}dx_{-},
\end{equation}
the diffeomorphism can be decomposed into two generators
\begin{equation}
	\xi=\xi^{(+)}+\xi^{(-)},
\end{equation}
with
\begin{equation}
	\begin{split}
		&\xi^{(+)}=V^+\partial_{+}-\frac{r}{2}\partial_{+}V^+\partial_{r}
		\\&\xi^{(-)}=V^-\partial_{-}-\frac{r}{2}\partial_{-}V^-\partial_{r},
	\end{split}
\end{equation}
where $(x^+,x^-)$ have the $2\pi$ period. Thus it can be expanded as the Fourier modes $V^{\pm}_{m}=e^{imx_{\pm}}$
\begin{equation}
	\xi_{m}=e^{imx}\partial_{x}-\frac{imr}{2}e^{imx}\partial_{r}.
\end{equation}
The communicators of the diffeomorphisms obey the Witt algebra
\begin{equation}
	i[\xi_{m},\xi_{n}]=(m-n)\xi_{m+n}.
\end{equation}
An important observation is that the Lorentz-violating parameter \(l\) neither appears in the generators nor alters the asymptotic symmetry algebra. Its sole effect is to rescale the \(g_{rr}\) component of the metric by a constant factor. Hence, the boundary CFT structure is left intact. This finding will prove crucial for interpreting the black hole entropy.

\subsection{Conserved Charges and Cardy Entropy Discrepancy}
We now turn to the computation of conserved charges and the construction of the quantum algebra for the spacetime under consideration. Rewriting the BTZ-like black hole in Fefferman--Graham coordinates, we obtain
\begin{equation}
	ds^2=(1+l)\frac{L^2dr^2}{r^2}-(rdx_{+}-\frac{L^2 H_{-}dx_{-}}{r})(rdx_{-}-\frac{L^2 H_{+}dx_{+}}{r}),
\end{equation}
where the parameters $H_{\pm}$ are defined as 
\begin{equation}
	H_{-}=\frac{j}{4L}+\frac{m}{4}~,~~H_{+}=-\frac{j}{4L}+\frac{m}{4}.
\end{equation}
Through CPSM, the conserved charges of the  diffeomorphisms $\xi^{(\pm)}$ are given by
\begin{equation}
	\delta\mathcal{L}_{m}^{\pm}=\frac{1}{16\pi}\int k_{\xi_{m}}[\delta g,g]=\frac{1}{8\pi}\int d\phi L\sqrt{1+l}e^{imx_{\pm}}\delta H_{\pm},
\end{equation}
which gives 
\begin{equation}
	\mathcal{L}_{m}^{\pm}=\frac{1}{8\pi}\int d\phi L\sqrt{1+l}e^{imx_{\pm}} H_{\pm}.
\end{equation}
We next examine the commutator of the conserved charges. The quantum algebra is then obtained by imposing the canonical commutation relations, which yield
\begin{equation}
	i\{\mathcal{L}_{m}^{\pm},\mathcal{L}_{m}^{\pm}\}=(m-n)\mathcal{L}_{m+n}^{\pm}+\frac{i}{16\pi}\int_{\partial\mathcal{M}}k_{\xi_{m}}[\delta_{\xi_{n}}g,g],
\end{equation}
we can obtain 
\begin{equation}
	\frac{i}{16\pi}\int_{\partial\mathcal{M}}k_{\xi_{m}}[\delta_{\xi_{n}}g,g]=\frac{Lm^3\sqrt{1+l}}{8}\delta_{m+n,0}
\end{equation}
The central charge can be directly read-off
\begin{equation}\label{c-btz}
	\frac{Lm^3\sqrt{1+l}}{8}\delta_{m+n,0}=\frac{cm^3}{12}\delta_{m+n,0}~,~~\rightarrow~c=\frac{3L\sqrt{1+l}}{2}.
\end{equation}
In the limit \(l \to 0\), our results consistently reduce to the pure AdS\(_3\) case of Einstein gravity. Furthermore, the linear term in \(m\) appearing in the Virasoro algebra can be obtained by a constant shift of the generators
\begin{equation}
	\tilde{\mathcal{L}}^{\pm}_{m}=\mathcal{L}^{\pm}_{m}+\frac{c}{24}\delta_{m,0} \,.
\end{equation}
As a consequence, the algebra of conserved charges can finally be written as
\begin{equation}
	i\{\tilde{\mathcal{L}}^{\pm}_{m},\tilde{\mathcal{L}}^{\pm}_{n}\}=(m-n)\tilde{\mathcal{L}}^{\pm}_{m+n}+\frac{c}{12}(m^2-1)m\delta_{m+n,0}.
\end{equation}
This establishes that the BTZ-like black hole admits a dual description in terms of a two-dimensional conformal field theory. Accordingly, by choosing the thermal field double (TFD) state as the microscopic configuration, we may identify the CFT thermal entropy with the black hole's microscopic entropy. The latter is computed via Cardy's formula \cite{Cardy:1986ie}, which for a 2D CFT at temperature \(T\) yields
\begin{equation}\label{SC-btz}
	S_{CFT}=2\pi(\sqrt{\frac{c_{L}E_{L}}{6}}+\sqrt{\frac{c_{R}E_{R}}{6}}),
\end{equation}
where $E_{L}$ and $E_{R}$ can be expressed as the zero-mode of the Virasoro generator
\begin{equation}\label{ELR}
	E_{L}=\tilde{\mathcal{L}}^{+}_{0}-\frac{c}{24}=\frac{L\sqrt{1+l}H_{+}}{4}~,~~ E_{R}=\tilde{\mathcal{L}}^{-}_{0}-\frac{c}{24}=\frac{L\sqrt{1+l}H_{-}}{4}.
\end{equation}
By substituting Eq(\ref{c-btz}) and Eq(\ref{ELR}) into Eq(\ref{SC-btz}), we can obtain the Cardy entropy
\begin{equation}
	S_{CFT}=\frac{\pi r_{0}}{2}\sqrt{1+l}.
\end{equation}
Upon comparing the Wald entropy,
\begin{equation}
	S_{\text{Wald}} = \frac{\pi r_{0}}{2}(1+l),
\end{equation}
with the microscopic entropy \(S_{\text{CFT}}\) obtained from the dual CFT via Cardy's formula, we observe a nontrivial discrepancy:
\begin{equation}
	S_{\text{Wald}} = \sqrt{1+l}\, S_{\text{CFT}}.
\end{equation}
This mismatch indicates that the Wald entropy does not coincide with the holographic counting of microstates.

This entropy mismatch constitutes the central result of this section and highlights a key distinction between our analysis and previous work \cite{Ding:2025cno}. In contrast to earlier approaches that modified the Cardy formula to enforce equality between the two entropies, we contend that such a modification is unjustified. Indeed, since the asymptotic symmetry group remains unchanged, the dual CFT must be identical to that of the standard BTZ black hole. Consequently, the discrepancy cannot be resolved by altering the CFT side; it must instead be attributed to the gravitational sector. While the three-dimensional BTZ case is somewhat degenerate due to the absence of bulk dynamical degrees of freedom, our result already signals that Lorentz violation produces a genuine modification of the relationship between macroscopic and microscopic entropy.

\section{An attempt to resolve the entropy discrepancy}
In this section, we discuss a possible approach to resolve this mismatch between Wald formalism entropy $S$ and Cardy entropy $S_{C}$.

Discrepancies in black hole entropy have also been reported in investigations of black hole thermodynamics within Horndeski gravity. A novel viewpoint on this topic was recently put forward in Ref. \cite{Hajian:2020dcq}. The authors revealed that the radial direction plays a distinctive role in Horndeski gravity. Since gravitational waves no longer travel at the speed of light along the radial direction, gravitons propagate in an effective metric obeying the modified dispersion relation \(g_{\mu\nu}^{(eff)}k^{\mu}k^{\nu}=0\). Accordingly, the surface gravity and temperature relevant to gravitons are those defined by this effective metric. They further argued that the black hole entropy derived from the symplectic two-form using the effective temperature represents the true physical entropy.

In our scenario, we examine the spacelike vector field given in Eq. (\ref{ansatz-bee}), which has only an r-component and depends solely on the radial coordinate r. This vector field shares the same structural properties as the scalar gradient \(b_{\mu} \sim \partial_{\mu}\phi(r)\) in Horndeski gravity. Accordingly, the matter sectors of Bumblebee gravity and Horndeski gravity can be viewed as  equivalent. Guided by the above framework, we calculate the effective temperature and the corresponding black hole entropy using the same approach.

Previous work in Ref.\cite{Liang:2022hxd} has comprehensively studied gravitational wave polarization in bumblebee gravity, where the Bumblebee field alters the corresponding dispersion relation. For the spacelike vector field considered here, the propagation speed \(c_g\) is derived as follows (see Refs.\cite{Maluf:2014dpa,Amarilo:2023wpn,Liang:2022hxd})
\begin{equation}\label{cg}
	c_{g}^2=
	\begin{cases}
		1, & \text{for gravitons moving normal to } b_{\mu}, \\
		1+\gamma b^2, & \text{for gravitons moving along } b_{\mu}.
	\end{cases}
\end{equation}
We then follow the procedure presented in Ref. \cite{Hajian:2020dcq} to derive the effective temperature. For gravitational waves propagating along \(k_{\mu}\), we have \(c_g = 1\) (equal to the speed of light) when \(k_{\mu}\) is orthogonal to \(b_{\mu}\), i.e., \(k_{\mu}b^{\mu}=0\). When \(k_{\mu}\) is parallel to \(b_{\mu}\), the effective metric can be directly extracted from the gravitational wave dispersion relation (\ref{cg}), as given below
\begin{equation}
	g_{\mu\nu}^{(\text{eff})} = g_{\mu\nu} - \frac{\gamma}{1+\gamma\, b^2} \, b_{\mu}b_{\nu}.
\end{equation}
In fact, the effective metric is defined only up to an overall constant factor, such that \(g_{\mu\nu}^{(\text{eff})} \to A(b_{\mu},g_{\mu\nu})\, g_{\mu\nu}^{(\text{eff})},\)where A is a function constructed from \(b_{\mu}\) and \(g_{\mu\nu}\). This global factor can generally be fixed via a suitable normalization condition. For example, when a timelike Killing vector \(\xi=\partial_{t}\) exists in Einstein gravity, it satisfies
\begin{equation}\label{xisq}
    \xi^2\Big|_{r\rightarrow\infty}=-1.
\end{equation}
In our case, the vacuum condition \(b_{\mu}b^{\mu}=\text{constant}\) imposed on the bumblebee field makes the factor A produce a constant shift at asymptotic infinity. For this reason, we still adopt (\ref{xisq}) to fix this factor, and we find that \(A=1\). Following the argument in Ref. \cite{Hajian:2020dcq}, we then derive the effective surface gravity via the standard method
\begin{equation}
	\kappa_{\text{eff}} = \sqrt{-\frac{1}{2}\,\nabla^{\mu}_{(\text{eff})} \xi^{\nu}\, \nabla_{\mu(\text{eff})} \xi_{\nu}}.
\end{equation}
where $\nabla_{\mu(eff)}$ is the covariant derivative operator compatible with the effective metric $g_{\mu\nu}^{(\text{eff})}$. It implies the temperature
\begin{equation}
    \tilde{T}=\frac{\kappa_{\text{eff}}}{2\pi}=\frac{h'(r_0)}{2\pi\sqrt{W(r_{0})}}.
\end{equation}

Turning now to our newly derived five-dimensional rotating black hole solution, the effective temperature is thus given by
\begin{equation}
    \tilde{T}=\frac{r_{+}^4-\nu^2}{2\pi r_{+}^3\sqrt{(\nu^2+r_{+}^4)}}.
\end{equation}
Compared with previous results, the two temperatures differ by a factor of \(\sqrt{1+l}\), namely
\begin{equation}
    \tilde{T}=\sqrt{1+l}T.
\end{equation}
Since the symplectic two-form on the horizon reads \(k_{\xi}[g,\delta g]|_{r_{h}}=\tilde{T} \delta \tilde{S}\), we extract the associated entropy using this effective temperature. The resulting entropy is
\begin{equation}
   \tilde{S}=\frac{\pi^2r_{0}\sqrt{\nu^2+r_{0}^4}}{2}\sqrt{1+l}.
\end{equation}
Then the first law is given by
\begin{equation}
    \delta M=\tilde{T} \delta \tilde{S}+\Omega_{+}\delta J.
\end{equation}
Following the procedure in the previous section, we adopt the Frolov–Thorne vacuum to describe the microstates of the extremal rotating black hole and construct its microscopic entropy.  It is worth to note that the Frolov–Thorne temperature is defined by a Boltzmann factor 
 \begin{equation}
     e^{-(\omega -m\Omega_{+})/T_{H}}=e^{-n_{L}/T_{L}+n_{R}/T_{R}}.
 \end{equation}
Now, the temperature is redefined as $T_{H}=\tilde{T}$. Correspondingly, the Frolov–Thorne temperatures now read
\begin{equation}
    \tilde T_{L}=\frac{h_{2}}{2\pi\omega'(r_{0})\sqrt{W_{0}}}=\frac{1}{2\pi}~,~~T_{R}=0.
\end{equation}
Substituting this expression into the Cardy formula yields the corresponding Cardy entropy
\begin{equation}\label{SC-kerr-1}
    \tilde S_{C}=\frac{\pi^2}{3}c_{L}\tilde T_{L}=\frac{\pi^2\nu^{3/2}}{\sqrt{2}}\sqrt{1+l}.
\end{equation}
It now exactly matches the thermal entropy of the corresponding extremal black hole
\begin{equation}
    \tilde{S}=\frac{\pi^2\nu^{3/2}}{\sqrt{2}}\sqrt{1+l}=\tilde S_{C}.
\end{equation}
Thus, the microscopic origin of black hole entropy is successfully explained. 

We can also revisit the BTZ black hole and find that its effective temperature is modified to
\begin{equation}
	\tilde{T}=\frac{\kappa_{eff}}{2\pi}=-\frac{j^2+4r_{0}^4\Lambda}{8\pi r_{0}^3}.
\end{equation}
Compared with Eq.~(\ref{T-btz}), the temperature is rescaled by \(\sqrt{1+l}\). From the Wald formalism, the new entropy is then given by
\begin{equation}
	\tilde{S}=\frac{\pi r_{0}}{2}\sqrt{1+l}.
\end{equation}
 The first law turns out to be
\begin{equation}
	\delta M=\tilde{T}\delta \tilde{S}+\Omega\delta J.
\end{equation}
On the other hand, applying the similar procedure as in Eq.~(\ref{SC-btz}) yields the Cardy entropy, which we find to be precisely equal to the thermal entropy
\begin{equation}
    \tilde{S}=\tilde S_{C}.
\end{equation}
Thus microscopic origin of black hole entropy in $D=3$ is also naturally resolved.

We now turn to a brief assessment of this approach. It is worth noting that, in other extended gravity theories such as Lovelock gravity, the gravitational wave dispersion relation is also modified—yet no corresponding redefinition of the black hole temperature is required. This suggests that the temperature redefinition adopted here is not a generic feature of modified gravity. We therefore view it as a possible resolution to the entropy discrepancy, but emphasize that it is not intended as a definitive prescription.

\end{document}